\documentstyle[12pt,epsfig]{article}
\def\np{\vspace{12pt} \noindent}

\begin{document}

\begin{center}
{\LARGE \bf {A correlation of the cosmic microwave sky with large scale
structure}}
\end{center}
\begin{center}
{\large \bf {Stephen Boughn$^1$ \& Robert Crittenden$^2$}}
\end{center}
\begin{center}
{ \bf {$^1$Haverford College, Haverford PA 19041 USA}} \\
{ \bf {$^2$Institute of Cosmology and Gravitation, Portsmouth PO1 2EG UK}} \\
\end{center}

{\np \bf
The fluctuations in the cosmic microwave background (CMB) have proved an invaluable tool for 
uncovering the nature of our universe.  The recent dramatic data provided by the 
WMAP satellite \cite{wmap} have confirmed previous indications that the expansion of the universe
may be accelerating\cite{bops}, driven by a 
cosmological constant or similar dark energy component.  One consequence 
of dark energy is the suppression of the rate of gravitational collapse of matter
at relatively recent times.  This causes fluctuations in the CMB to be created as the 
photons pass through nearby large scale structures, a phenomenon known as the integrated
Sachs-Wolfe (ISW) effect.  The result is additional  
large scale fluctuations in the CMB which are correlated with 
the relatively nearby (i.e., at redshift $z \sim 1$) matter distribution \cite{ct}.  
Here we report evidence of 
correlations between the WMAP data and two all sky probes of large scale structure, the hard X-ray
background observed by the HEAO-1 satellite \cite{heao} and the NVSS survey of radio galaxies \cite{nvss}.
Both observed correlations are consistent with an ISW origin, indicating that we are seeing the 
impact of dark energy on the growth of structure.   
}

\np
In the standard model of the origin of structure, most of the fluctuations were imprinted on the CMB
at the epoch of last scattering, when the universe was 400,000 years old ($z\simeq 1100.$) 
The ISW effect induces extra fluctuations only when matter domination ends and the dark 
energy becomes important dynamically ($z\sim 1$.)  When this happens, the gravitational 
potentials of large, diffuse concentrations and rarefactions of matter begin to decay and 
the energy of photons passing through them changes by an amount that depends on the depth of the 
potentials.  The amplitude of these ISW fluctuations tends to be small 
compared to the fluctuations originating at the epoch of last scattering except on very large scales. 
However, since ISW fluctuations were created more recently, it is expected that the CMB fluctuations 
should be partially correlated with tracers of the large scale matter distribution, e.g., with the
distribution of distant galaxies.

\np 
Detecting the relatively weak correlation of the CMB with the distribution of galaxies
requires nearly full sky surveys out to redshifts $z\sim 1$.  Focus has thus has been on 
luminous active galaxies, which are believed to trace the mass distribution on large scales.  
While active galaxies emit at a wide range of frequencies, the most useful maps are in 
the hard X-rays (2-10 KeV), where they dominate the X-ray sky, and in 
the radio, where the number counts are dominated by sources at $z\sim 1$.  
The full sky map of the intensity of the hard X-ray background (XRB) made by the HEAO-1 
satellite \cite{heao} and a map of the number density of radio sources provided by the NRAO VLA sky 
survey (NVSS) \cite{nvss} are the two best maps to search for this effect.
In order to predict the expected level of the ISW effect in these two surveys, 
it is essential to know both the inherent clustering of the
sources and their distribution in redshift.  The former can be determined from
previous studies of the X-ray and radio auto-correlation functions \cite{bc,bck} while the latter
can be estimated from deep, pointed surveys\cite{cowie,dp}.
The detailed considerations for calculating the amplitude of the 
ISW effect can be found elsewhere\cite{ct, bct}.   

\np 
Previous attempts at measuring the correlation by cross correlating the CMB
with both the HEAO-1 and the NVSS maps have yielded only upper limits \cite{bc,bck}. 
Here we repeat these analyses 
with the WMAP satellite data.  We use two CMB maps generated from the WMAP
data: the ``internal linear combination'' (ILC) map \cite{wmap2};  and the ``cleaned'' map of Tegmark,
de Oliveira-Costa and Hamilton \cite{tegmark}. 
The dominant ``noise'' to the cross correlation signal is due to the fluctuations in the CMB itself and is well
characterized, while the instrument noise is negligible on the angular scales relevant to the present analysis.
To reduce possible residual Galaxy and nearby radio source 
contamination, these maps were masked with the most aggressive mask provided by the WMAP 
team\cite{wmap2}. 
The X-ray map was similarly masked and, in addition, was corrected for 
several large scale systematics including a linear drift of the detectors \cite{bck}.  Finally, the NVSS
catalog was corrected for a systematic variation of source counts with declination \cite{bc}.
The sky coverage was $68\%$ for the CMB maps, $56\%$ for the NVSS map, and $33\%$ for the X-ray map.
The aggressive masking of the X-ray maps was chosen because of the low resolution of the map ($3^{\circ}$)
and the larger number of foreground sources. 
Significantly, however, our results are relatively insensitive to the 
corrections and level of masking of the maps.

\begin{figure}[ht]
\begin{center}
\epsfig{file=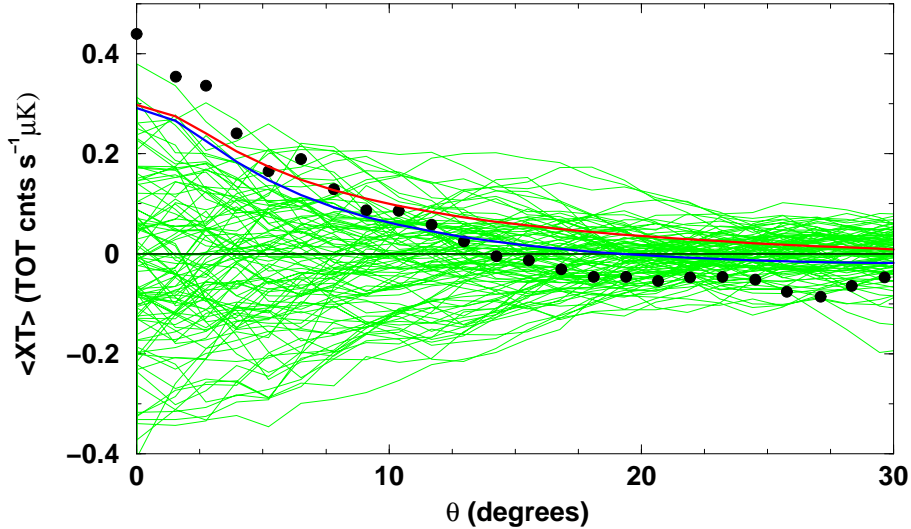,width=12cm,angle=0}
\caption{
{\bf The X-ray intensity measured by HEAO-A1 is correlated with the microwave
sky measured by WMAP at a higher level than would be expected by chance correlations.}
Here we plot the cross correlation
between the X-ray intensity fluctuations and the CMB temperature fluctuations
along with the theoretical predictions for the ISW effect in a cosmological constant
($\Omega_\Lambda = 0.72$), the
best fit WMAP model for scale invariant fluctuations.  To give an idea of the level of accidental
correlations, the green curves show the result of correlating the X-ray map with 100
independent Monte Carlo realized CMB maps with the same power spectrum as the WMAP data.
The variance increases at smaller angular separations, where there are fewer pairs of pixels contributing to the
correlation and one can see that the signals in neighboring bins are highly correlated for a given realization.
Due to the shape of the expected correlation, the signal to noise is greatest at smaller angular separations.
For $\theta = 0^{\circ}$,
$1.3^{\circ}$, and $2.6^{\circ}$, the Monte Carlo trials exceed the amplitude of the actual X-ray/CMB correlation only
$0.3$\%, $0.8\%$, and $0.3\%$ of the time respectively.  These correspond to $2.4$ to $2.8~\sigma$.
At larger angular separations, the observed correlations appear to fall faster than predicted by theory.
The blue line shows the theoretical predictions if the quadrupole and octupole modes are suppressed
as suggested by the measured WMAP temperature spectrum.
While it seems to
fit the data better, the larger angular separations have very low signal to noise.} 

\label{xt.fig}
\end{center}
\end{figure}

\begin{figure}[ht]
\begin{center}
\epsfig{file=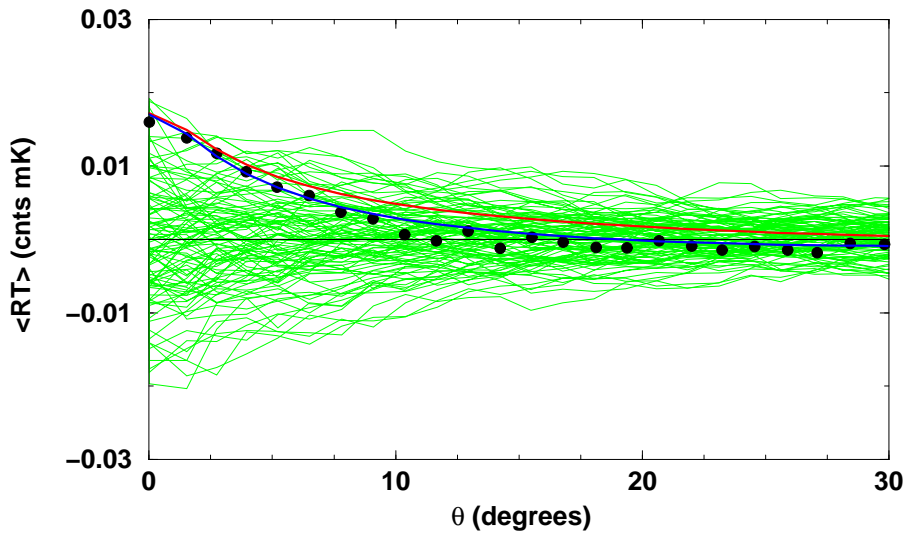,width=12cm,angle=0}
\caption{
{\bf The NVSS radio galaxies also appear correlated with the microwave
sky, but at lower confidence level than the X-rays.} Here we plot the correlation between the
radio galaxy number counts and the WMAP temperature maps.  The other curves are as in Figure \ref{xt.fig}.
The Monte Carlo trials exceed the amplitude of the actual radio/CMB correlation in the lowest three bins
$1.2\%$, $1.9\%$, and $3.4\%$ of the time respectively, corresponding to a $1.8$ to $2.3~\sigma$ signal
detection.  Again, there is good agreement with the theoretical predictions, with the signal falling off
faster than predicted at larger angles (at fairly low statistical significance.)
The consistency of the NVSS and X-ray CCF's suggests that the signal
is not the result of unknown systematics in
either the X-ray or the NVSS map.}
\label{rt.fig}
\end{center}
\end{figure}

\np
The maps were expressed as 24576 $1.3^\circ \times 1.3^\circ$ 
pixels in an equatorial quadrilaterized spherical cube projection on 
the sky \cite{quad}.
The cross correlation function (CCF) of the X-ray and CMB maps was computed according to
\begin{equation}
CCF(\theta) = {1 \over N_{\theta}} \sum_{i,j} (I_i-\bar{I}) (T_j-\bar{T})
\end{equation}
where the sum is over all pairs of pixels, $i,j$, with angular separation $\theta$, $N_{\theta}$
is the number of pairs of pixels separated by $\theta$, $I_i$ is the X-ray intensity of the 
$i^{th}$ pixel, $T_j$ is the temperature of the $j^{th}$ pixel of the the CMB map, and \
$\bar{I}$ and $\bar{T}$ are the mean values of these quantities.  The CCF of the NVSS and CMB
maps is similarly defined.  Figures \ref{xt.fig}, \ref{rt.fig} are plots of the CCF's for angular separations 
from 0 to 30 degrees along with the expected ISW signal.  The CCF's using the ILC and ``cleaned'' CMB maps were 
entirely consistent with each other as expected  and we plot the averages
of the results for the two maps.  Four hundred Monte Carlo 
simulations of the CMB sky were also cross correlated with the actual X-ray and radio maps.  The $rms$ values
of these trials are taken as the errors for each point.  
These errors are highly correlated due to the large scale structure in the maps. 
For comparison, the errors were also computed 
from the data themselves by rotating the maps with respect to each other so that for small angular 
separations, the pixel pairs were far removed from one another on the sky.  The two error estimates
were consistent.
The significance of the detection of the ISW effect can be estimated by fitting the amplitude of the 
theoretical profile to the data.  Minimum $\chi ^2$ fits 
indicate that the significance of the signal is roughly $2.0-2.5\,\sigma$ for the radio/CMB CCF 
and $2.5-3.0\,\sigma$  for the X-ray/CMB CCF, depending on the number of bins included in the fits.   
These values are consistent with number of times the  CCF($\theta$)
of the Monte Carlo trials exceed the values of the measured CCF($\theta$) for the first few data points where
the ISW signal is largest.

\np 
The possibility that the signal is arising from unmasked foreground sources is unlikely since the observed 
correlations do not change when the Galactic cuts are varied nor when the aggressiveness of
the masking is changed.  It is significant that the amplitudes of the CCF's are 
essentially unchanged when no masking at all is applied to the CMB maps.  It is also unlikely
that a small number of diffuse foreground sources is responsible for the signal.  When the
different hemispheres of the maps are analyzed separately the results are consistent. 
The fluctuations caused by inverse Compton scattering of CMB photons (as they travel through
the ionized intergalactic medium of rich clusters of galaxies), the Sunyaev-Zeldovich effect,
is expected to dominate only at much smaller angles and be anti-correlated with the matter distribution\cite{ps}.  
Finally, it is possible that microwave emission from the radio/X-ray sources themselves could
result in the positive correlation of the maps.  However, extrapolations of the frequency spectra
of the radio galaxies indicate that the microwave emission is much smaller than the observed
signal \cite{bc}.  In addition, the clustering of radio sources is on a much smaller angular scale than 
the apparent signals, and any such contamination would appear only at 
$\theta=0^\circ$ and not at larger angles\cite{b98}.

\np 
While the detection of ISW type signals in the two CCF's is only at the $2$ to $3~\sigma$ level, we find it 
significant that both these signals are consistent with the theoretical predictions with no free parameters.
However, it should be emphasized that these
two measurements are by no means independent.  Both of them use the same CMB maps and, furthermore, the
X-ray background is highly correlated with the NVSS radio sources\cite{b98}.  
If the distribution of sources of one of the maps by chance coincided with the fluctuations of the CMB,
one would expect the other map also to be correlated with the CMB to some degree. 
The coincidence of the expected amplitudes of the two
signals is encouraging but by no means definitive since the signal to noise is not large.  On the other hand,
the detection of the ISW type signal in both CCF's gives a strong indication that they are not due to some 
unknown systematic effects in the maps.  The radio and X-ray data were 
gathered by quite distinct methods and it would be surprising if unknown systematics in the two maps 
were correlated in any way. We will present more detailed analysis of all these issues elsewhere. 

\np 
We tentatively conclude that we have observed the integrated Sachs-Wolfe effect. 
If so, these observations offer the first direct glimpse into
the production of CMB fluctuations and provide important, independent confirmation of the
standard cosmological model, dominated by dark energy.
It is worth noting that the power spectrum of CMB fluctuations  
does not show evidence of increased power on large angular scales ($\theta > 20^{\circ}$) as predicted by 
the ISW effect, but rather indicates that there is a 
significant deficit of power at these angular scales\cite{wmap}.
Our observed CCF's also appear to drop off somewhat faster in $\theta$ than
do the theoretical predictions indicating perhaps a deficit of large scale power in 
the CMB fluctuations produced locally. 
However, for $\theta > 8^{\circ}$ both the signal and theoretical prediction
are less than $1~\sigma$ so quantitative assessments are problematic.  We also note that the WMAP-NVSS
results have been independently analyzed by the WMAP team.

\np
{\small{We are grateful to Mike Nolta, Lyman Page and the rest of the WMAP team, as well 
as Neil Turok, Bruce Partridge and Bruce Bassett,   
for useful conversations.  RC acknowledges financial support from a PPARC fellowship.}}

\np
{\bf {Communicating author: Robert Crittenden, Robert.Crittenden@port.ac.uk}}
\pagebreak

\newpage

\begin{figure}[ht]
\begin{center}
\epsfig{file=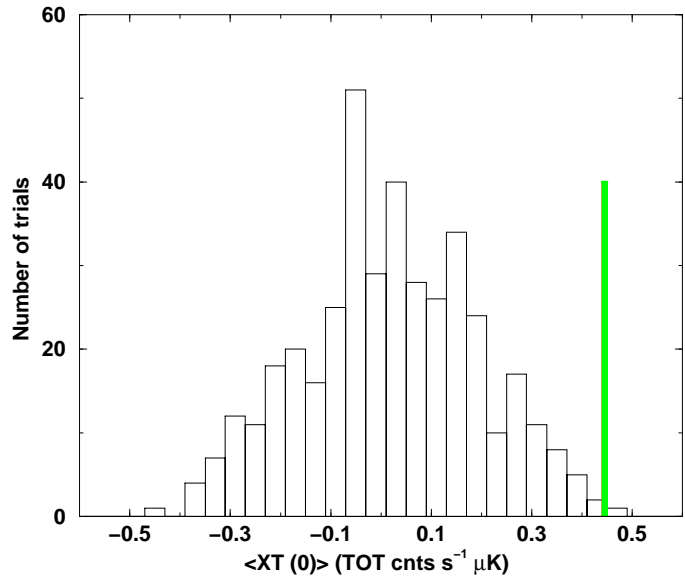,width=9cm,angle=0}
\caption{
{\bf The observed X-ray/CMB cross correlation for $\theta = 0^{\circ}$ (in green) exceeds the value found
for most Monte Carlo simulations.}
Here we plot the distribution of the correlation from 400 simulations, where the observed X-ray map
was cross correlated with random CMB maps with the same power spectrum as the observed WMAP CMB sky.
The underlying distribution of the Monte Carlos is not precisely Gaussian, but should be nearly Gaussian by
the central limit theorem.}
\label{hist.fig}
\end{center}
\end{figure}

\newpage 

\end{document}